# Academic information retrieval using citation clusters: In-depth evaluation based on systematic reviews

Juan Pablo Bascur[1, 2]: https://orcid.org/0000-0002-4077-1024

Suzan Verberne[2]: https://orcid.org/0000-0002-9609-9505

Nees Jan van Eck[1]: https://orcid.org/0000-0001-8448-4521

Ludo Waltman[1]: https://orcid.org/0000-0001-8249-1752

1 Centre for Science and Technology Studies, Leiden University, The Netherlands

2 Leiden Institute for Advanced Computer Science, Leiden University, The Netherlands

j.p.bascur.cifuentes@cwts.leidenuniv.nl

## Abstract
The field of science mapping has shown the power of citation-based clusters for literature analysis, yet this technique has barely been used for information retrieval tasks. This work evaluates the performance of citation-based clusters for information retrieval tasks. We simulated a search process with a tree hierarchy of clusters and a cluster selection algorithm. We evaluated the task of finding the relevant documents for 25 systematic reviews. Our evaluation considered several trade-offs between recall and precision for the cluster selection. We also replicated the Boolean queries self-reported by the systematic reviews to serve as a reference. We found that citation-based clusters' search performance is highly variable and unpredictable, that the clusters work best for users that prefer recall over precision at a ratio between 2 and 8, and that the clusters are able to complement query-based search by finding additional relevant documents.

## 1 Introduction
Researchers and other knowledge workers need special information retrieval (IR) tools because their IR tasks and practices differ from the general public and from each other (Ellis, 1993; Kuhlthau, 1991; Russell-Rose et al., 2018). Academic literature search is an essential part of any research project, and the most commonly used IR method is query-based retrieval: search using keyword queries to retrieve a ranked list of documents. However, some users complement this method with citation-based IR methods that follow the citations of the documents (Hemminger et al., 2007; Ortuño et al., 2013). These methods have two major advantages over query-based retrieval: 1) They are independent of the keywords, helping with lack of vocabulary knowledge or semantic ambiguity, and 2) they use the intellectual information of the citations, helping find documents that other researchers already connected. However, these methods can be timewise inefficient for users (Wright et al., 2014).

Given the prominence of citation clusters in scientometric research (Waltman & van Eck, 2012), it is remarkable that citation cluster-based IR (CCIR) is largely absent from the toolset of users (Wolfram, 2015). CCIR combines citation-based IR and cluster-based IR by making use of clusters of documents identified based on citation links. CCIR could allow users to also use approaches developed in scientometric research, such as science maps (Chen, 2017), cluster labeling (Sjögårde et al., 2021), and visualization software (van Eck & Waltman, 2010). CCIR offers two potential benefits over other citation-based IR methods: 1) it is less hindered by documents that cite the relevant literature poorly (Robinson et al., 2014) and 2) it communicates the topic structure of a document corpus, including the relative size of different topics and the relations between topics (Pirolli et al., 1996).

Effective cluster-based IR requires the clusters to group together the documents that are relevant for the IR task of the user (i.e., the cluster hypothesis (van Rijsbergen, 1979)). The extent to which this condition is fulfilled by CCIR is an open question. The answer may be different for different types of IR tasks (Hearst & Pedersen, 1996) and for different CCIR implementations. We consider one specific IR task, namely performing a literature search to write a systematic review (SR), and one specific CCIR implementation, namely a tree hierarchy of citation-based clusters of MEDLINE documents. As discussed below, we believe this to be a sensible use of CCIR. Moreover, data for experimentation was relatively easily available for this task. To determine the extent to which CCIR groups together relevant documents, we address the following research questions:

- What types of users are best served by CCIR?
- What types of SRs are best served by CCIR?
- What are the strengths and weaknesses of CCIR?

We answer these questions by simulating a CCIR search process, evaluating its performance and analyzing its results. We simulated the CCIR search process in the tree hierarchy with an algorithm that aims to simulate the behavior of a human user. The idea of a CCIR hierarchy is based on classical cluster-based IR strategies (Cutting et al., 1992; Jardine & van Rijsbergen, 1971) and on a frequently used scientometric approach for creating classification systems of science (Waltman & van Eck, 2012). We evaluated the performance of CCIR for the task of finding the relevant documents for 25 SRs from a benchmark dataset (Scells et al., 2017), using as performance reference the SRs' self-reported Boolean query search retrieved documents, obtained through intensive manual annotation. This task is well-suited for cluster-based IR because all relevant documents are considered equally important; the task is considered a Boolean retrieval task, so there is no ranking of documents. From these results we analyzed the different preferences of hypothetical users regarding the trade-off between precision and recall, the overlap between documents retrieved by CCIR and by a Boolean query, and how the topic of a SR affects its task performance.

To our knowledge, our work is the first study that evaluates the performance of CCIR. We additionally provide two outputs that can be reused by other researchers: 1) an evaluation protocol for clusters-based IR methods that uses SRs, and 2) an extension of the original SR dataset with the annotated Boolean queries.

This paper is organized as follows. We discuss related work in Section 2, explain out methodology in Section 3, show our results in Section 4, discuss our results in Section 5, and conclude our work in Section 6.

# 2 Related work

## 2.1 Science mapping

Our research on CCIR is part of a bigger trend of research that attempts to connect the fields of scientometrics and information retrieval. Experts agree that these fields have much to gain from each other (Frommholz et al., 2021; Mayr & Scharnhorst, 2015). While research on CCIR seems to have slowed down in recent years, research on clustering methods in the field of scientometrics continues to move forward.

Closest to our research are the citation clusters used for science mapping and field delineation studies (Chen, 2017; Cobo et al., 2011). It has been shown that these clusters create communities of documents with semantic similarity (i.e., a common topic) (Klavans & Boyack, 2006) and that they provide insights for analyzing these documents (Small & Garfield, 1985). Citation clusters are also used to represent communities of documents in the visualization of a citation network (which is a network of documents and their citations to each other) (Chen, 2006; van Eck & Waltman, 2017). Text similarity-based clusters, both on their own (Callon et al., 1983) and enriched with citations (Ahlgren et al., 2020; Janssens et al., 2008), have also been used to map science. Waltman et al. (2020) compare citation-based similarity clusters with text similarity-based clusters. We decided not to include the use of text similarity in our research because text similarity-based cluster IR is already a well-studied method (see Section 2.3).

## 2.2 Citation-based IR

Citation-based IR methods are frequently used in academic search. The most common method is to retrieve the documents that cite or are cited by a given document (a.k.a. citation tracking). A further step of this method is to track the citations of these retrieved documents (a.k.a. snowballing). Some of the developments in citation-based IR are tools to track citations (Chandra et al., 2021; *Inciteful*, 2022; Janssens et al., 2020; Liang et al., 2011; Madeira & Vot, 2018; Pitt et al., 2022; van Eck & Waltman, 2014), protocols to find relevant documents to write a SR by tracking citations (Belter, 2016; Horsley et al., 2011), tools that delineate fields by tracking citations (Zitt, 2015), methods to rank search results by tracking citations (Belter, 2017; Mutschke & Mayr, 2015), and methods to find the seminal documents of a topic by tracking citations (Haunschild & Marx, 2020). Additionally, citation-based IR is addressed by the communities around the workshop series Bibliometric-enhanced Information Retrieval (BIR) (Frommholz et al., 2021) and the related workshop series Bibliometric-enhanced Information Retrieval and Natural Language Processing for Digital Libraries (BIRNDL) (Cabanac et al., 2017).

The most significant difference between CCIR and citation tracking is that CCIR creates clusters and retrieves documents using the structure of the whole citation network, while citation tracking retrieves documents using only the structure of the documents closest to the initially selected document in the citation network. Both methods focus on different aspects of the citation network, so both can be valuable to the academic IR toolset.

## 2.3 Cluster-based IR

Cluster-based IR methods retrieve one or more clusters of documents, and these clusters are usually based on text similarity. These methods have been used for academic search both in commercial (*Iris.Ai*, 2019) context and academic (*Open Knowledge Maps: A Visual Interface to the World's Scientific Knowledge.*, 2019) contexts, and have also included the text from cited documents in their similarity score (Abbasi & Frommholz, 2015). Non-academic IR has also been used to cluster web search results (Stefanowski & Weiss, 2003). Additionally, the seminal Scatter/Gather browsing model (Cutting et al., 1992) (on which we draw inspiration for our evaluation) proposes a user interaction protocol where the user removes irrelevant documents over several iterations by creating new sets of documents using the clusters from the previous iteration. Bascur et al. (2019) proposed specifications for a CCIR tool that uses the Scatter/Gather model.

Cluster-based IR works have a wide methodological variety, reflected in the following methodological choices:

- Relatedness attribute between documents: Connections (e.g. citations, as we did) or shared elements (e.g. text, authors, keywords);
- Which set of documents to cluster: Either the whole corpus (as we did) or a subset of the corpus that is retrieved by a query;
- What is the structure of the clustering solution: Either hierarchical (as we did) or flat (a.k.a. independent clusters);
- How to select clusters during the evaluation: Either select clusters using knowledge of the document relevance (as we did) or select clusters using a query match;
- How to retrieve documents during the evaluation: Either retrieve all documents within a cluster (as we did) or retrieve only some.

Our purpose is not to compare the pros and cons of each of these methodological choices. Instead, our focus is on evaluating the specific methodological choices considered in our work. Similar to our work is the work of He et al. (2019), who visualize academic search results using, among other elements, citation-based clusters. The difference between their approach and ours is that we use the clusters as a means to retrieve documents, while they use the clusters for visualization of search results. In their work, they showed that their visualization can increase the efficiency (i.e., completion time) and user satisfaction for complex tasks, but not for simple tasks. This result suggests that the effectiveness of CCIR may depend on the task. Therefore, we look at individual SR tasks to see how the effectiveness differs between them.

Measuring the effectiveness of clustering, both for IR and for other purposes, is not trivial, as no clustering solution can satisfy every possible search task (Yuan et al., 2022). Our approach is to measure clustering effectiveness without the participation of real users (a.k.a. offline evaluation). Many other studies have adopted the same approach. For instance, Abdelhaq et al. (2013) created a metric for evaluating Twitter data clustering based on the stability and coverage of the most common keywords in a cluster. In a bioinformatics example, Atkinson et al. (2009) evaluated the effectiveness of a gene similarity network clustering by observing to what extent each cluster had a single gene function. Yuan et al. (2022) created novel metrics that consider the number of clusters necessary to retrieve a given percentage of the relevant documents. De Vries et al. (2012) created an evaluation framework where the relevant documents are known and the clustering solution is compared with a random baseline. Abbasi and Frommholz (2015) evaluated clustering with a simulation where a virtual user already knows which are the relevant documents. Our evaluation is most similar to the latter two studies because our cluster selection algorithm already knows which are the relevant documents, which is a common assumption in evaluation of retrieval methods (Manning et al., 2008).

## 3 Method

### 3.1 Task design and data collection

The task we address is to find the documents necessary to write a given SR. The data that we use for this task comes from the dataset published by Scells et al. (2017) (from now on referred to as the Scells dataset). This dataset contains:

- 177 SRs published by the Cochrane library between 2014 and 2016.
- The references of each SR that belong to the included studies or excluded studies category of that SR. We consider both categories necessary for the task of writing a SR, so we included documents from both categories in the set of relevant documents of the task (see below for an explanation).
- The self-reported Boolean query that the authors of each SR used when they searched using the OVID search platform with the MEDLINE database, hereafter referred to as the Boolean query.

We intend to retrieve the documents that the authors of the SR found in their search, thus we use the authors' Boolean queries to retrieve documents. We retrieved these documents following these steps:

1. We manually confirmed that the Boolean queries in the Scells dataset were the same as the ones self-reported by the SRs, and when this was not the case, we used the self-reported one.
2. We translated the Boolean queries from the OVID format into the PubMed format because the OVID search platform does not have an API service, while the PubMed search platform does (*PubMed API*, 2018) and it also includes the MEDLINE database. We translated the formats using the TRANSMUTE software (Scells et al., 2018) and then we manually checked that the translation was correct (i.e., that both formats would retrieve the same documents). Some translations were not possible because the OVID search platform provides functionalities that the PubMed search platform does not (e.g., word distance-based arguments). A full report on the translations and how we handled difficult cases can be found in the supplementary material, Tables S1 and S2.
3. For each SR, we performed a search using the PubMed API based on the PubMed Boolean query, and we included the retrieved documents in the document set retrieved by the Boolean query.
4. We removed from the retrieved document set the documents that were not in the citation network (which is described in Section 3.2). We also removed from the relevant document set (see below) the documents that were absent from the document set retrieved by the Boolean query in order to maintain consistency between both sets (i.e., so that the relevant document set is a subset of the document set retrieved by the Boolean query).

To improve the quality of our evaluation, we selected a subset of the SRs in the Scells dataset to be used in our evaluation. Our selection criteria were:

- The relevant document set contains at least 10 documents. We chose this value because with fewer relevant documents, the increase in recall for each retrieved document would be more than 0.1 and we wish a more fine-grained increase to facilitate interpretation of the results.
- The number of retrieved documents self-reported by the authors (i.e., from all their search sources) is of a similar order of magnitude (i.e., between 10 times less and 10 times more) as the size of the document set retrieved by us with the Boolean query. This condition excludes SRs whose self-reported number of retrieved documents is vastly different from ours.

This selection resulted in 25 SRs (see Figure 4A in Section 4 for the number of relevant documents per SR), of which 7 were published in 2014, 10 in 2015 and 8 in 2016. The number of SRs may seem small, for instance in comparison with the work by Janssens et al. (2020), who used 250 SRs. However, we manually annotated the Boolean queries, which is very labor intensive. Additionally, while the number of SRs is modest, the number of document in our citation networks is very large (~7 million per network, see below).

Cochrane library SRs have, for our purposes, three categories of documents in their references:

- Included studies: Studies that provide information that advances the objective of the SR.
- Excluded studies: Studies that were considered for the included studies category but were discarded because they did not match the selection criteria of the SR.
- Additional references: Documents that were not considered for the included studies category.

The Cochrane library has a clear rule for which documents should go into the excluded studies category: When a user discards a document, after they have read the document full text to any extent, the document is an excluded study, else it is not (e.g., discarded after reading the abstract).

We decided to regard the excluded studies as relevant documents for the retrieval task because, by the above rule, the user needs to find and read these documents in order to exclude them. Additionally, the selection criteria that discard an excluded study can be so particular (e.g., number of participants in the study) that we believe it is not reasonable to expect an IR tool to be able to discard these documents.

### 3.2 Citation network

We needed to create a citation network for the tree hierarchy of clusters. We used the in-house Dimensions database, which contains all the documents included in MEDLINE and also their citation links. We created the citation network following these steps:

1. We retrieved all the documents contained in the Dimensions database.
2. We removed all the documents published the same year or later than the SRs to make sure we do not provide unfair advantageous information to the clustering (see below). Therefore, we created a different citation network for each publication year in the Scells dataset: One until 2013, another until 2014 and another until 2015.
3. We limited the documents of the citation networks to the ones available in the MEDLINE database, because the self-reported Boolean queries were performed exclusively within the MEDLINE database. We identified the MEDLINE documents using the PubMed database available at Leiden University's Centre for Science and Technology Studies (CWTS).
4. Because of the computing resources needed to handle large citation networks, we limited the publishing years of each network to 11 years (2003-2013, 2004-2014, and 2005-2015).

The sizes of the citation networks were:

- Citation network 2003-2013: 6,549,426 documents, 81,284,099 citation links.
- Citation network 2004-2014: 6,879,646 documents, 86,001,142 citation links.
- Citation network 2005-2015: 7,194,514 documents, 90,164,417 citation links.

Documents that are in the reference lists of a given SR are connected to the SR by a citation link. These connections help the clustering algorithm to put all these documents in the same cluster, which would artificially increase the performance of CCIR. This is not fair because in a real scenario these connections could not exist because the SR has not been published yet. We removed not only these connections, but all the documents published in the same year and in later years because they could be influenced by these connections. Because we remove the documents published in the same year, we may also remove some documents that existed before the publication of the SR. However, none of the relevant documents were removed in this process.

### 3.3 Simulation of CCIR

In this section we explain how we simulated the CCIR search process so we can evaluate the performance of CCIR.

#### 3.3.1 Clustering

We created a tree hierarchy of clusters for each citation network. We started by clustering the documents into at most 10 clusters, based on the idea that in practice it may be difficult for users to handle more than 10 clusters. Then, the documents of each cluster were again clustered into at most 10 smaller clusters, and so on. As discussed below, the documents that could not be included in these clusters were excluded from the tree. This process created a nested tree of clusters with a depth of 13 levels (not counting the root level). We only clustered into smaller clusters the clusters that contained relevant documents because otherwise they were irrelevant for the evaluation.

We performed the clustering using a methodology built on the work of Waltman and van Eck (2012). This methodology is used in combination with the Leiden algorithm (Traag et al., 2019). This combination provides a state-of-the-art approach for document clustering in the field of scientometrics. This approach has been used in a large number of research articles (e.g. Boyack et al., 2020; Held & Velden, 2022; Sjögårde & Ahlgren, 2020). It is also used in products of the analytics companies Elsevier (Elsevier, n.d.) and Clarivate (Potter, 2020). We therefore consider it the state-of-the-art approach for citation-based clustering.

In the methodology of Waltman and van Eck (2012), the tree hierarchy is built in a bottom-up manner while we take a top-down approach. We made this change because it reflects how a real user would create a tree, going from the general to the specific. It also saves computer resources by not creating sub-clusters for clusters that are of no interest. Another change is that Waltman and van Eck merged small clusters based on a cluster size threshold, while we merged small clusters based on a number of clusters threshold (at most 10 clusters, as mentioned before). We made this change because for a real user it is more intuitive to control the maximum number of clusters than the minimum number of documents per cluster.

The purpose of the Leiden algorithm is to assign documents to clusters based on the connections between the documents. The algorithm rewards pairs of documents in the same cluster that are connected by a citation link and penalizes pairs of documents in the same cluster that are not connected. The magnitude of the penalty is determined by the resolution parameter of the algorithm, which must be provided externally. A higher resolution leads to more and smaller clusters.

Mathematically, the clustering algorithm maximizes the following quality function:

$$V(x_1, \ldots, x_n) = \sum_i \sum_j \delta(x_i, x_j)(a_{ij} - r) \qquad (1)$$

In this quality function, i and j are documents, $x_i$ is the cluster of document i, and r is the resolution parameter. $a_{ij}$ equals 1 if there is a citation link between documents i and j. Otherwise $a_{ij}$ equals 0. $\delta$ equals 1 if $x_i$ and $x_j$ are equal (i.e., documents i and j are in the same cluster). Otherwise $\delta$ equals 0.

The Leiden algorithm returns the clustering solution that maximizes Equation 1. To limit the number of clusters per clustering (i.e., children clusters per parent cluster) to at most 10, we merged the smaller clusters following these steps:

1. If there are more than 10 clusters in the clustering solution, select the smallest cluster. If there is a tie in the size, randomly select one of the smallest clusters. If the number of clusters is 10 or fewer, stop.
2. If there are no citation links between the documents in the selected cluster and documents outside the selected cluster, remove the selected cluster from the clustering solution and then go back to step 1.
3. For each cluster other than the selected cluster, calculate the highest resolution under which this cluster would merge with the selected cluster (method from Waltman and van Eck (2012)). This resolution is always lower than the current resolution because otherwise the clustering algorithm would have already merged these clusters.
4. Merge the selected cluster with the cluster for which the highest resolution was obtained in step 3, and then go back to step 1.

The resolution parameter must be provided externally, but the literature has not yet established a rule of thumb for selecting a suitable value (although the work of Sjögårde and Ahlgren (2018, 2020) goes in that direction). We therefore used our own heuristic. Using a trial-and-error approach, we tried to find resolution values for each level so that the following conditions were satisfied as much as possible:

- The size of the 10 largest clusters after merging was similar to the size of these clusters before merging. This condition aims to minimize the effect of cluster merging.
- The 10 largest clusters after merging were of similar size. This condition aims to avoid creating one or a few clusters with a disproportionally large number of documents.

Our heuristic resulted in a resolution of $2\times10^{-5}$ for the first level of the tree hierarchy. For each subsequent level we multiplied the resolution by 3. At level 13 the resolution is greater than 1 ($2 * 10^{-5} * 3^{12} = 1.06$) which is why we have 13 levels (a resolution greater than 1 yields only singleton clusters).

### 3.3.2 Cluster selection

We use a greedy algorithm to select the clusters, starting from the root of the tree hierarchy. The algorithm goes down the tree hierarchy selecting child clusters based on their score, until none of the child clusters has a score higher than the currently selected cluster (see Figure 1). We use a greedy algorithm because this reflects how a real user would navigate a tree hierarchy. The score function is the F-score of retrieving the documents in a cluster, determined based on the relevant documents of a given SR:

$$F_\beta = (1 + \beta^2) \times \frac{precision \times recall}{(\beta^2 \times precision) + recall} \quad (2)$$

The precision and recall of each cluster are calculated based on the number of documents in the cluster (i.e., number of positives), the number of relevant documents in the cluster (i.e., number of true positives), and the number of relevant documents not in the cluster (i.e., number of false negatives). A real user does not have access to these numbers. The greedy algorithm therefore simulates an optimistic scenario in which a user is able to accurately assess the quality of different clusters.

The parameter β of the F-score function (Equation 2) reflects how a hypothetical user balances recall against precision (van Rijsbergen, 1979): Lower values of β favor precision, while higher values favor recall. If β = 1, precision and recall have equal weight. For each SR we retrieve several clusters, each one using different values of β to cover a wide range of precision-recall trade-offs: β ∈ {0.125, 0.25, 0.5, 1, 2, 4, 8, 16, 32, 64, 128}. The idea of using a greedy algorithm and different values of β to reflect real users is inspired by the "what-if" experiments methodology (Azzopardi et al., 2011).

### 3.4 Quantitative analysis

For our quantitative evaluation, we group the results of the SRs according to value of β used by the cluster selection algorithm. In this way, we can compare the aggregated results for different values of β. We report the number of retrieved documents, the tree-level of the retrieved cluster, precision, recall, and F-score (β = β used by the cluster selection algorithm).

We report four more metrics that are generated by comparing the cluster selection algorithm results with the Boolean query retrieved documents:

- Intersection proportion of the cluster selection algorithm: Proportion of the documents retrieved by the cluster selection algorithm that are also retrieved by the Boolean query.
- Intersection proportion of the Boolean query: Proportion of the documents retrieved by the Boolean query that are also retrieved by the cluster selection algorithm.
- Ratio of retrieved documents: Number of documents retrieved by the cluster selection algorithm divided by the number documents retrieved by the Boolean query.
- F-score difference: F-score of the cluster selection algorithm minus the F-score of the Boolean query ($\beta = \beta$ used by the cluster selection algorithm).

The purpose of the F-score difference is to evaluate the performance of CCIR while also taking into consideration the difficulty of the task for the authors of the SR. We refrain from using the F-score difference to make claims about the relative performance of CCIR compared to the Boolean query. We do not consider such claims to be justified, because there are too many issues that we are not able to take into account in our analyses. For instance, we assume that the Boolean query retrieves all relevant documents, but we are unable to assess the accuracy of this assumption. Also, in practice, a Boolean query is written over several iterations of trial and error. We are unable to analyze the impact of this iterative process, since we have access only to the final version of a Boolean query.

Instead of directly comparing the performance of a CCIR approach with a Boolean query approach, our quantitative analysis focuses on answering the following questions:

- To what extent does the performance of CCIR varies between individual SRs? We answer this by analyzing the dispersion of the F-score difference grouped by of $\beta$.
- How similar are the sets of documents retrieved by CCIR and the Boolean query? We answer this by analyzing the intersection proportion of both CCIR and the Boolean query
- For which values of $\beta$ is CCIR more effective? We answer this by analyzing most of the quantitative metrics, and how their values change when the value of $\beta$ increases or decreases.

### 3.5 Qualitative analysis

In our qualitative analysis we address the following questions:

- How does the nature of a SR affect the performance of CCIR and a Boolean query?
- What type of documents does CCIR or a Boolean query retrieve or miss?

We address these questions by an expert reading of the SRs performed by the first author of our paper (Juan Pablo Bascur), who is trained in the biomedical field, and supported by an expert in Boolean query searches for biomedical purposes (Jan W. Schoones).

We performed the qualitative analysis on the retrieved documents of three SRs. We selected the SRs based on their F-score difference for $\beta = 4$ (we used $\beta = 4$ because it had the highest recall dispersion, which helps highlight the differences between SRs; see Section 4). We selected the SRs with the lowest, highest and third highest F-score difference, which in the Scells dataset correspond to the ids SR59 (Peinemann et al., 2013), SR47 (Cousins et al., 2016) and SR80 (Ma, 2015), respectively.

For each SR, we characterized:

- Goal: The question that the authors of the SR want to answer.
- Needs: The nature of the documents that the authors need to retrieve to achieve the goal.
- Boolean query components: The components of which the Boolean query consist. A component is a group of Boolean terms that belong to the same topic.

For each SR we also selected one of the clusters that CCIR retrieved for this SR, that we subjectively found it had good precision and recall (hereafter known as the optimal cluster). We also selected from the clusters that CCIR retrieved the parent and the child of the optimal cluster to expand the range of our analysis, but we discarded the child clusters because they were so small that they did not provide qualitative information. Therefore, we selected the parent of the optimal cluster, hereafter known as the parent cluster.

We inferred the topic of each set of documents (these are, the clusters and the document retrieved by the Boolean query) from the titles of the documents. For the bigger document sets, we facilitated this process by inferring the topics from the most common noun-phrases in the titles of the documents. We extracted noun phrases from titles using the spaCy Python library (Honnibal et al., 2020).

To guide our analysis, we use Venn diagrams of the overlap between the relevant documents, the selected clusters of CCIR and the documents retrieved by the Boolean query. We also look for documents retrieved by CCIR but not by the Boolean query that, given their nature, could have been relevant documents if the authors of the SR had found them.

## 4 Results

### 4.1 Quantitative results

In this section we describe the quantitative analysis of the 25 SRs evaluation results. Figure 2 shows the precision, recall, F-score and F-score difference, Figure 3 shows the intersection proportions, Figure 4 shows the number and ratio of retrieved documents, and Figure 5 shows the level of the selected clusters.

#### 4.1.1 To what extent does the performance of CCIR vary between individual SRs?

Figure 2E shows that the F-score difference values have a large dispersion: within $\beta$ groups the interquartile range is 0.2 or higher, and the highest range (at $\beta = 4$) is 0.5. This result shows that the performance varies between SRs, and it highlights the importance of analyzing individual SRs in the qualitative analysis presented in Section 4.2.

#### 4.1.2 How similar are the sets of documents retrieved by CCIR and the Boolean query?

Figure 3 shows that these two sets of documents are very different because their intersection proportion is very low. We analyzed Figure 3 focusing on three $\beta$ groups, which we selected based on Figure 4D: when both document sets are of the same size ($\beta = 16$), when the CCIR set is 10 times bigger than the Boolean query set ($\beta = 128$), and when the CCIR set is 10 times smaller than the Boolean query set ($\beta = 2$). When both sets are the same size and when the CCIR set is 10 times bigger, the intersection proportion is surprisingly low: 0.1 for the former (Figures 3A and 3B) and 0.5 for the latter (Figure 3B). When the CCIR set is 10 times smaller, the proportion is also low (0.6), but additionally this value starts to fall dramatically on the subsequent groups of $\beta$ (Figure 3A).

### 4.1.3 For which values of β is CCIR more effective?

Figure 5 shows that the tree-level of the selected clusters is linearly correlated with the value of β (using our powers of 2 scale), or in other words, the median level goes up by 1 level for each sequential β value.

Figure 4D shows that, for β between 2 and 128, the CCIR retrieved document set was between 10 times smaller and 10 times bigger than the Boolean query document set. Figure 2B shows that the β groups after β = 8 have less precision than the Boolean query (0.025, Figure 2A). Figure 2C shows that recall improves little after β = 8. Therefore, we think that the results of groups β = 2, β = 4 and β = 8 balance size, precision and recall the best. Also, outside these groups the balance decreases much faster from β = 1 to the lower values of β than from β = 16 to the higher values of β.

## 4.2 Qualitative results

In this section we describe the qualitative analysis of three selected SRs and their evaluation results. Figure 6 shows their Venn diagram of the intersection between the Boolean query, the CCIR and the relevant documents. Table 1 shows their quantitative data, Table 2 shows their characterization and Table 3 shows the topic of their sets of documents. The details on the construction of their Boolean query components can be found in supplementary material Figures S1, S2 and S3, and their topics in supplementary material Tables S3, S4 and S5.

### 4.2.1 SR59: Retinoic acid post consolidation therapy for high-risk neuroblastoma patients treated with autologous hematopoietic stem cell transplantation

This SR had the lowest F-score difference and also a high Boolean query precision (Table 1). Its goal was to determine if patients with the condition _Neuroblastoma_ recuperate better from the treatments _Chemotherapy_ and _Bone Marrow Transplant_ if they are treated with the medication _Retinoic Acid_ (Table 2).

The document set of Boolean query and the two clusters had similar topics, but the cluster topics were missing the component _Retinoic Acid_ (Table 3), which is one of the needs of SR59 (Table 2). This suggests that CCIR did not create a cluster with _Retinoic Acid_, and we wonder why. All the relevant documents of SR59 clearly share a common topic (we read their titles) so it would seem that they should be mostly in the same CCIR cluster. An explanation for this mystery seems to be given by the topic of the parent cluster. Here, we found that the topic fulfills the needs of SR59, except that instead of _Retinoic Acid_ it has the component _131L-MIBG_, which is a medication with similar uses to _Retinoic Acid_. It seems then that the existence of a cluster with the needs of SR59 and _Retinoic Acid_ was mutually exclusive with the existence of a cluster with the needs of SR59 and _131L-MIBG_, and CCIR created the latter instead of the former because of its higher fitness. This likely resulted in CCIR spreading the relevant documents of SR59 among other clusters, decreasing the F-score difference value.

The Boolean query of SR59 is missing the component _Bone Marrow Transplant_ from the needs of SR59 (Table 2), yet the Boolean query achieves a high precision (Table 1). This is because the combination of the components _Neuroblastoma_ and _Retinoic Acid_ was so infrequent in the literature that it was enough for Boolean query. This shows that the Boolean query can give high precision for highly specific needs.

### 4.2.2 SR47: Surgery for the resolution of symptoms in malignant bowel obstruction in advanced gynaecological and gastrointestinal cancer

This SR had the highest F-score difference (Table 1). Its goal was to determine how effective the treatment *Surgery* is to treat the condition *Intestinal Obstruction* when caused by the conditions *Gynecological Cancer* or *Gastrointestinal Cancer* (Table 2).

We could not identify the topic of the Boolean query document set because the most common noun-phrases were present in only a minor portion of the documents. This could be either because the set of documents was big and therefore has too much diversity, or because it has several disconnected topics, and we believe the latter explanation is the correct one. On the other hand, the topics of the two clusters (Table 3) were similar to the needs of SR47 (Table 2).

We believe that the Boolean query has several disconnected topics because the needs SR47 were hard to express in a Boolean query format, which ends up retrieving a noisy set of documents. The needs are documents on *Surgery* to treat *Intestinal Obstruction* due to *Gynaecological and Gastrointestinal Cancer* (Table 2). However, the Boolean query cannot specify if *Surgery* treats *Intestinal Obstruction* or treats *Gynaecological and Gastrointestinal Cancer*. This case shows that CCIR can help with searches where the relation between the Boolean query terms is ambiguous.

Additionally, we saw an interesting phenomenon happening with the topics of the clusters. Among their documents, there were three synonym noun-phrases that refer to intestinal obstruction: *Malignant Bowel Obstruction*, *Malignant Colorectal Obstruction* and *Malignant Colonic Obstruction*. The optimal cluster only had the first form, while the parent cluster had all three of them. This implies that the documents with the first form cite each much more intensely than the documents with the other two forms. We see no science-related reason for this to be the case, so we imagine that this citation pattern arises from a community of researchers with the same writing conventions that cite each other. This citation pattern shows one of the risks of CCIR and of citation-based clustering in general: The citations may not only represent an intellectual relationship between two documents, but also other non-scientific relationships that are of no use for IR purposes.

We saw another interesting phenomenon happening with the topics of the clusters. Two of the most common noun-phrases of the optimal cluster were _Inoperable Bowel Obstruction_ and _Octreotide_ (which is a medication for inoperable tumors). Both noun-phrases imply that their documents lack surgery, but _Surgery_ is a need of SR47. This shows that, even when the F-score difference value is high, CCIR may still not have created a cluster with the topic that the user needs.

### 4.2.3 SR80: Rituximab for rheumatoid arthritis (Review)

This SR had the third highest F-score difference (Table 1). Its goal was to evaluate the medication _Rituximab_ to treat the condition _Rheumatoid Arthritis_. There are two things we must mention for our analysis of SR80: First, that the medication _Rituximab_ belongs to a group of medications called _DMARDs_ (which means Disease-Modifying Antirheumatic Drugs), and second, that the needs of SR80 include comparing _Rituximab_ treatments with either no treatment (a.k.a. placebo) or other _DMARDs_ treatments (Table 2).

The topic of the Boolean query and the clusters is the same and fits the needs of SR80. This shows that CCIR created a cluster for the right topic. However, CCIR still missed several relevant documents, which shows that creating a cluster for the right topic can be insufficient. We believe that the reason these relevant documents were not in the clusters is that, even if two documents are about the same topic, they may be poorly connected to each other by direct or indirect citations due to the citing practices of their research community. This result challenges one of the core assumptions of CCIR: That two given documents that share a topic will be directly or indirectly well connected by citations.

It seems that the authors of the SR made the conscious decision of building the Boolean query in such a way that it sacrifices precision in favor of recall. This is suggested by the following difference between the required needs of SR80 and the Boolean query components of SR80 (Table 2): SR80 requires comparisons between treatments with _Rituximab_ (itself a _DMRAD_) and treatments with placebo or other _DMRADs_, but the Boolean query components do not require a document to mention _Rituximab_, resulting in several retrieved documents that do not serve the needs. We believe that the authors made this decision because they expected many documents that use _Rituximab_ to mention it in their metadata under the more general term _DMRADs_. This case shows that CCIR can help with searches where the Boolean query cannot be sufficiently specific.

An interesting observation is that, among the most common noun-phrases, the Boolean query mentions the same _DMRADs_ as the parent cluster, but the latter also mentions one extra _DMRAD_ (_Certolizumab Pegol_). This is interesting because the component _DMARDs_ of the Boolean query searched for all the available _DMARDs_, so it should also have found _Certolizumab Pegol_. We found that this happens because of the MeSH term that the component _DMARDs_ uses (_"Antibodies, Monoclonal"[Mesh Terms:noexp]_) does not retrieve _Certolizumab Pegol_ (which goes under _"Antibodies, Monoclonal, Humanized"[Mesh Terms:noexp]_). Biologically speaking, _DRMADs_ is better described by the latter MeSH term than by the former, but it seems that the convention of the National Library of Medicine is to use the former MeSH term for all _DRMADs_ except for _Certolizumab Pegol_. The authors may not have been aware of this because otherwise they presumably would have incorporated the second MeSH term in the Boolean query. We believe that this case shows that CCIR can help Boolean query users to ensure they include all necessary vocabulary in their Boolean query.

We wondered if any of the documents of the parent cluster with *Certolizumab Pegol* in their title may have been a relevant document if the authors of SR80 had seen the document during their literature search. We tested this hypothesis by comparing these documents with the needs SR80. We found one document (Weinblatt et al., 2012) which cannot be discarded based only on the title or the abstract, and therefore is a relevant document. This case shows that CCIR can find relevant documents that the Boolean query does not.

# 5 Discussion

In this section we discuss our findings in relation to our research questions and then discuss the limitations of our work.

## 5.1 What types of users are best served by CCIR?

We can answer questions about users by connecting user preferences for recall and precision with the $\beta$ value (user prefer recall $\beta$ times as much as precision). We saw that $\beta = 2$, $\beta = 4$ and $\beta = 8$ had the best balance, and that outside these $\beta$ values the balance decreases faster for lower $\beta$ values than for higher $\beta$ values. Therefore, we can say that CCIR serves best users that prefer recall over precision with a ratio between 2 and 8 times, and for users outside that range it serves higher ratios better than lower ratios.

We wondered if users that perform a literature search for a SR are within this range of ratios, and we used the Boolean queries values as a proxy to answer this. Figure 2A shows that the precision of the Boolean queries is between 0.01 and 0.06, and by definition the Boolean queries have a recall 1.0, so the ratio of recall over precision is 1 over 0.01-0.06, or 17-100, very far from our prior range of 2-8. While it is true that the recall of the Boolean query is unrealistically high, the recall would have to be 10 times lower for the ratio to be within the range, which, given that SR literature searches aim for maximum recall, is unlikely. Therefore, we believe that the users that are best served by CCIR are not users that do a literature search for SR. It is beyond our knowledge which type of user might prefer the range 2-8.

We saw that the median tree-level is sensitive to the $\beta$ value. While we do not have a standard to evaluate which levels are better for users, we know that the more a user prefers recall, the closer to the root, the less effort the user needs to make to reach that level.

We also saw that the Boolean query and CCIR retrieve different documents (Figure 3), and these documents could be relevant (analysis of SR80). Therefore, CCIR could serve users willing to use more than one IR method by finding more relevant documents.

## 5.2 What types of SRs are best served by CCIR?

We saw that there is a substantial variance among the F-score difference values of the SRs (Figure 2E), meaning that for some SRs, CCIR performs much worse than for others. We would imagine that, for CCIR, a SR with general needs (e.g. a disease) would perform better than a SR with specific needs (e.g. interaction between two medications), while the opposite would be true for Boolean queries (Carmel et al. (2006) analyzed how the needs affect query difficulty). However, the three SRs that we analyzed had specific needs (Table 2) yet one had bad performance and two had good performance. The only clue that we can use to infer the performance of a SR is in SR47: its need is hard to write as a Boolean query, so we can infer that IR methods not based on a Boolean query are likely to have an advantage. However, this inference is more about the bad performance of the Boolean query than the good performance of the CCIR.

## 5.3 What are the strengths and weaknesses of CCIR?

### 5.3.1 Strengths

CCIR may find documents that the Boolean query does not. We know this from the results of intersection proportions (Figure 3), where it shows that CCIR and the Boolean query retrieve different documents. We also know this from the newly discovered relevant document of SR80.

CCIR may reduce the noise of searches that are hard to write as a Boolean query. We know this from how CCIR performs well for SR47 and SR80: The former's Boolean query could not be sufficiently specific because the Boolean query format does not allow to specify subject-object relations between terms. The latter's Boolean query could not be specific because of the risk of missing documents with poorly annotated metadata.

CCIR may help expand the vocabulary used in a Boolean query. We know this from our experience with SR80. By looking at the difference between the noun-phrases of the parent cluster and the Boolean query of SR80, we realized that the Boolean query was missing a relevant search term which was likely not considered by the authors of the Boolean query.

### 5.3.2 Weaknesses

CCIR may not create a cluster with the exact topic that the user needs. We know this because in SR47 and SR59 there was a divergence between the user needs and the topic of the CCIR sets of retrieved documents. The tree hierarchy did not had a cluster with the same topic as the user needs, which may happen because documents may relate to multiple topics.

The performance for a given SR can be unpredictable. We know this because of the high dispersion of the F-score difference values (Figure 2E) and because the characteristics of SR59, SR47 and SR80 did not give a clue about their performance.

Documents that share the same topic may be poorly directly or indirectly connected in a citation network. We know this from our experience with SR80. While a cluster with the relevant topic was retrieved, several relevant documents were missing. Also, the noun-phrases differences between the retrieved documents of the optimal cluster and the parent cluster of SR47 suggest that the optimal cluster was created based on the citation practices of the authors instead of the topic of the documents. Potentially, this issue could be diminished by combining citation-based and semantic-based clustering.

The clusters at the highest levels have too many documents, which makes the topic of the clusters hard to interpret for a real user because the documents are so diverse. This is a serious problem because selecting the wrong cluster at this level is a critical mistake (Willett, 1988). Our evaluation did not suffer from this issue because CCIR already knows in which clusters the relevant documents can be found. In a real situation, a user may be able to handle this issue if they know at least some of the relevant documents, and then they could even select clusters bottom-up instead of top-down (Van Rijsbergen & Croft, 1975). Alternatively, the user can create the tree hierarchy with fewer documents.

## 5.4 Limitations of this work

We identified four potential limitations to our work.

First, we did not cover all the possible clustering solutions. We used a single clustering solution, instead of using several clustering solutions or letting a user create clustering solutions on the run. Some of the characteristics of the tree hierarchy could have been different, like the clustering algorithm that we used, the clustering resolution parameters, the number of child clusters, the number of levels and the fact that we created the tree hierarchy by a top-down division of clusters instead of a bottom-up agglomeration of clusters.

Second, we did not cover all the possible citation networks. We used a citation network of direct citations, and not a more densely connected citation network using co-citations (Small, 1973) or bibliographic coupling (Martyn, 1964), which when combined with direct citation improve the representation of the structure of science (Waltman et al., 2020). We made the citation network using the full corpus, but we could also have used for example the documents retrieved from a query, which some studies reported to be more effective for cluster-based IR (Tombros et al., 2002).

Third, the cluster selection algorithm does not reflect fully realistic (and noisy) user behavior. The cluster selection algorithm knows the relevant documents – an assumption commonly made in information retrieval evaluation –, which a real user would not. A real user would have to select the clusters based on their own personal evaluation of which cluster is more likely to contain the relevant documents, and also they would have to evaluate when to stop going down the tree hierarchy. This process would take cognitive effort, which our evaluation does not consider. A less cognitively heavy alternative for a user could be to eliminate a cluster that does not contain relevant documents and then create a new clustering solution, as there is likely to be an obvious candidate for elimination. This is the same process as selecting more than one cluster, as we discussed in the weaknesses (Section 5.3.2), and we decided against implementing it in the evaluation because it would create too many steps and the clustering would take too much computational resources. Another unrealistic behavior is that the cluster selection algorithm never chooses the wrong cluster, unlike a real user. We could have implemented mistakes by giving imperfect information to the cluster selection algorithm, but we decided not to so to have less variables that could affect the interpretation of our results. Finally, it is not realistic to allow the cluster selection algorithm to choose very small clusters (size between 1 and 10 documents) because this size of clusters does not appear in real situations (as discussed by Willet (Willett, 1988)). Future work could address more noisy user behavior, similar to user behavior modeling in information retrieval (Hofmann et al., 2013).

The final limitation is that it could be argued that the Boolean queries we used are not realistic. A real Boolean query is created over several iterations, where the creators of the query keep refining the query until they are satisfied with the search results. Our evaluation does not consider this. Also, our Boolean queries had a recall of 1.0 (i.e., they found all the relevant documents), which is unlikely for a real IR method. Additionally, we only considered the documents retrieved by the Boolean query on MEDLINE, while the authors of the SRs usually used more than one database or method to search for documents, including the expert knowledge of their colleagues. We did not include more sources because it would be too much effort to retrieve the documents of each method and to harmonize the results between SRs that used different methods. Finally, the translation from OVID format to PubMed format is likely to have modified the set of retrieved documents, especially if the Boolean query used OVID-specific features (like distance between words). We tried to remove the cases with the biggest modification of the set of retrieved documents by removing the SRs with Boolean queries that retrieved a number of documents too different from the number documents self-reported by the authors (see Section 3.1).

# 6 Conclusion

In this work we have shown some of the advantages and limitations of using CCIR for academic search, both for generic CCIR and for our specific tree hierarchy implementation. We have also introduced an evaluation protocol for cluster-based IR methods with the task of finding relevant documents for SRs. This protocol can be used and modified by other researchers. We release our data for use by other researchers in the form of the three tree hierarchies, the set of relevant documents and the set of documents retrieved by the Boolean query, the latter one created through intensive manual annotation. The current CCIR implementation can be used as a straightforward CCIR tool of value for real users.

Our research shows that the best served users are those who prefer recall over precision 2 to 8 times. Users that prefer even more recall, like SR users, are less well served, and users that prefer more precision are the worst served. CCIR may complement Boolean query searches in various ways: it may help SR users that have problems to state their requirements as Boolean queries, it may suggest terms for Boolean queries, and it may retrieve relevant documents not retrieved by a Boolean query.

A problematic aspect of CCIR is that performance varies significantly because there sometimes is no cluster that contains the topic of the SR. This may happen because documents may relate to multiple topics, leading to clusters that do not match with the topic of the SR. It may also happen because of a lack of citation connections between the documents related to the topic of interest. Another problematic aspect is that the current implementation of CCIR demands a high cognitive effort from a user.

For future work related to CCIR, interesting research directions are how to improve its performance (how to create better clusters, re-clustering based on the selection of multiple clusters by a user, mixing with semantic-based clustering), how it compares to other IR methods (especially citation-based or cluster-based methods) and how real users interact with it (how to select clusters, how to complement with other IR tools).

# Acknowledgements

We would like to thank Jan W. Schoones for his expert support in biomedical Boolean queries, and Vincent Traag and Roel van der Ploeg for their invaluable feedback. We are also grateful to an anonymous reviewer for their comments on our work.

## Competing interests

We have no competing interests to declare.

## Funding information

We did not receive any funding for the research presented in this paper.

## Data availability

The code used to run the experiments in this paper is available in GitHub (https://github.com/jpbascur/citation_clusters_evaluation) and the data is available in Zenodo (Bascur, 2022).

# Figures and tables
## Figures

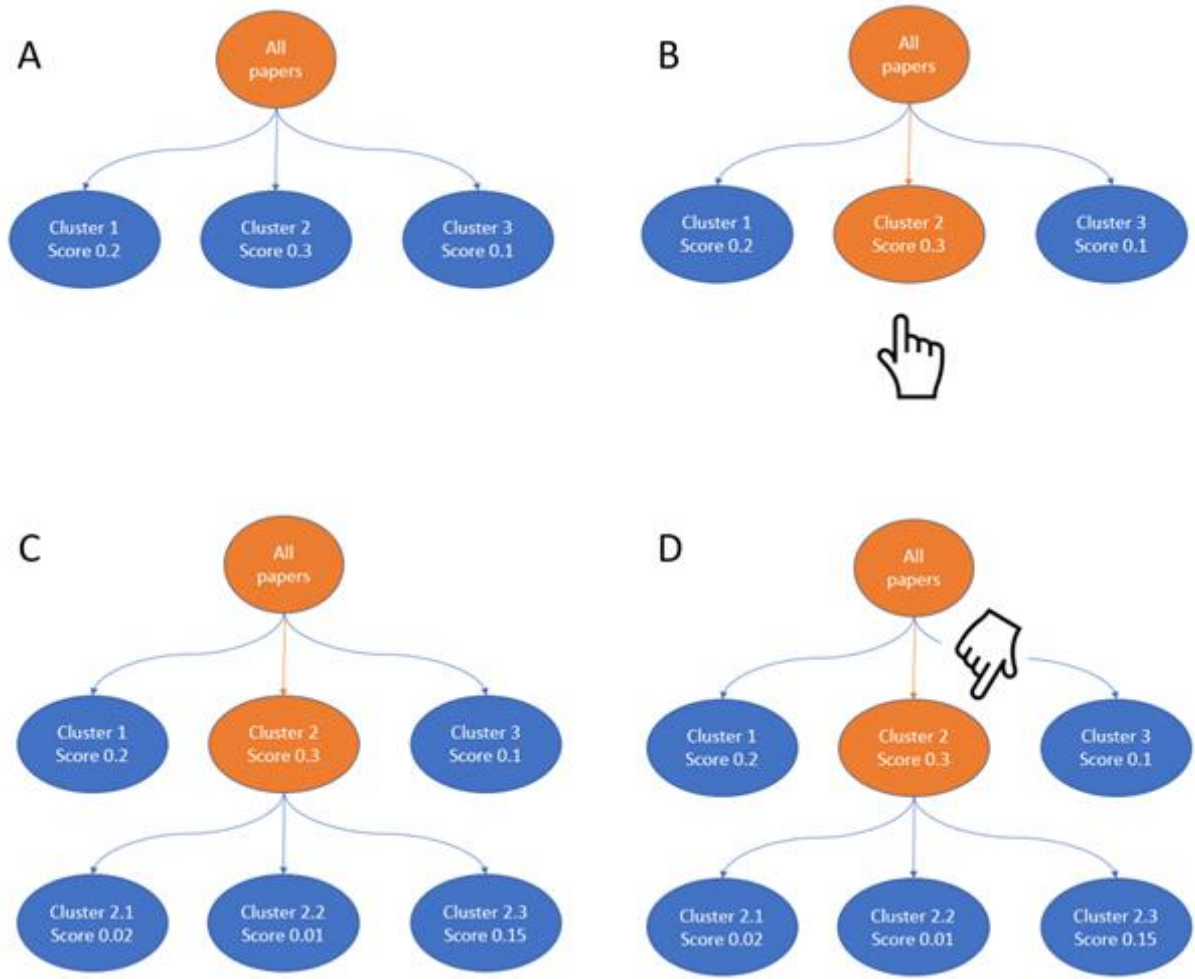

**Figure 1. Cluster selection algorithm.** The bubbles represent clusters of documents. The text in a bubble shows the label and the score of a cluster. The lines are the connections between the parent and the child clusters in the tree hierarchy. The arrows point toward the child clusters. Only the child clusters of the selected clusters are shown. The orange bubbles represent the clusters selected by the algorithm. The orange lines indicate the path followed by the algorithm. The pointer finger shows the selection of the algorithm. **A**: Calculate the score of each cluster at the highest level of the tree hierarchy (Clusters 1, 2, and 3). **B**: Select the cluster with the highest score (Cluster 2). **C**. Calculate the score of each child cluster of the selected cluster (Clusters 2.1, 2.2, and 2.3). **D**. Retrieve the cluster that was already selected (Cluster 2) because it has a higher score than any of the child clusters.

A
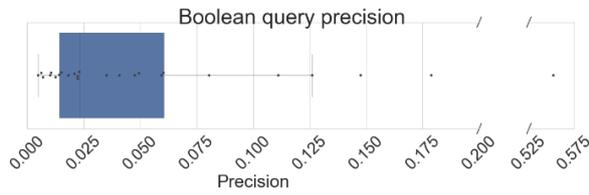

B
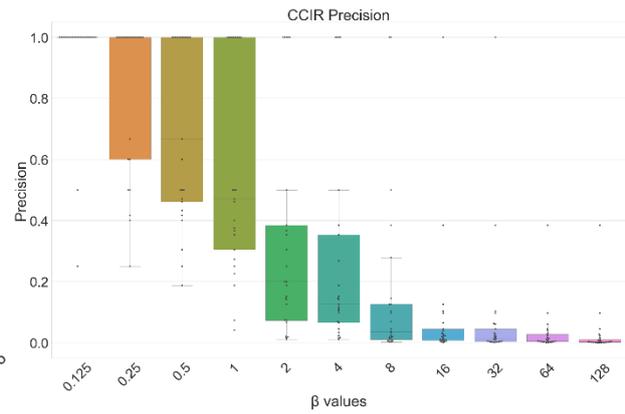

C
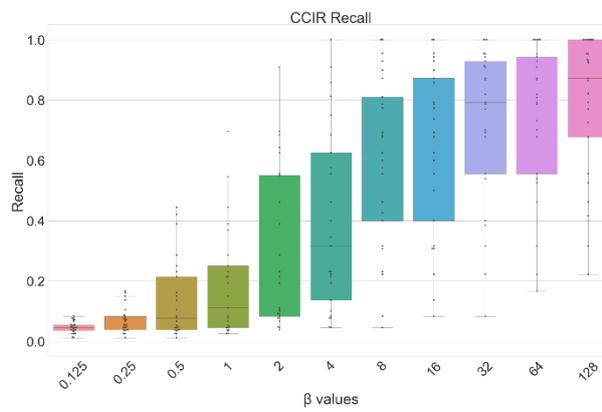

D
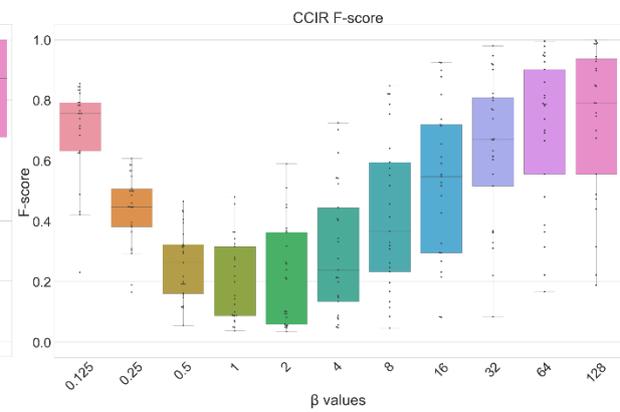

E
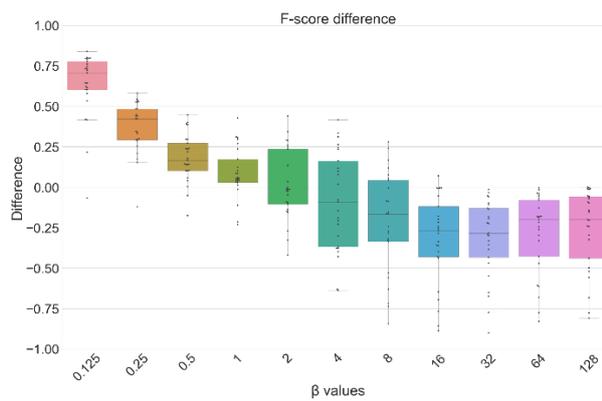

F
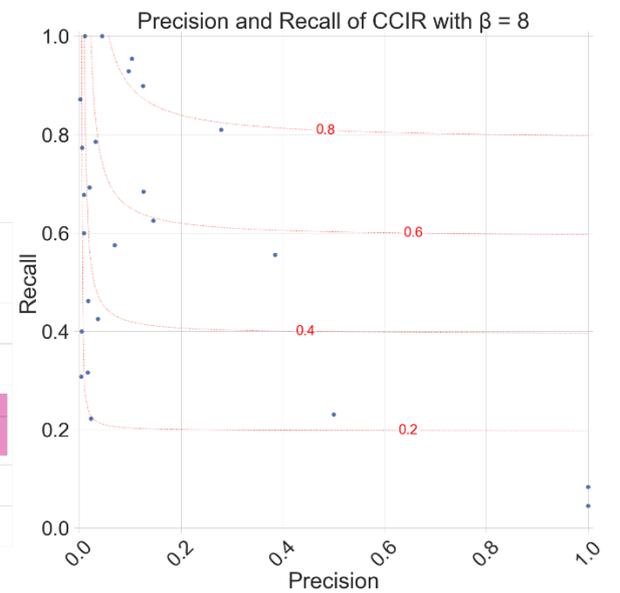

**Figure 2. Precision, Recall and F-Score. A**: Precision of the Boolean query. Each data point is a SR, and the X axis is the precision. **B to E**: Each data point is a SR, the X axis is the β group, and the Y axis is the respective metric of that β group for that SR. **B**: Precision of CCIR. **C**: Recall of CCIR. **D**: F-Score of CCIR. **E**: F-score difference between CCIR and the Boolean query (CCIR minus Boolean query). **F:** Precision and recall of β = 8. Each data point is a SR, the X axis is the precision of CCIR, the Y axis is the recall of CCIR, and the red lines are the isocurves of the F-score (β = 8).

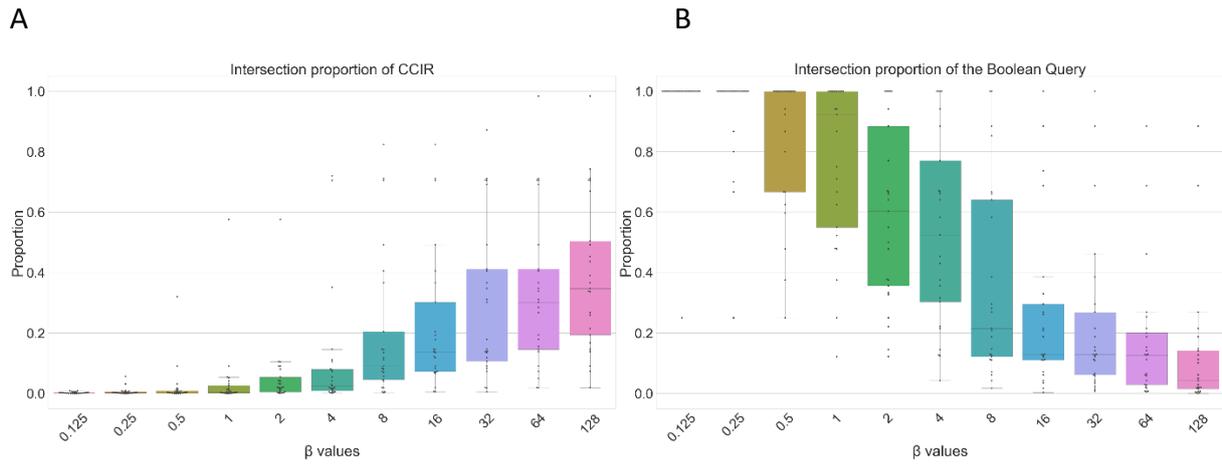

**Figure 3. Intersection proportions. A and B**: Each data point is a SR, the X axis is the β group, and the Y axis is the respective metric of that β group for that SR. **A**: Intersection proportion of CCIR. **B**: Intersection proportion of the Boolean query.

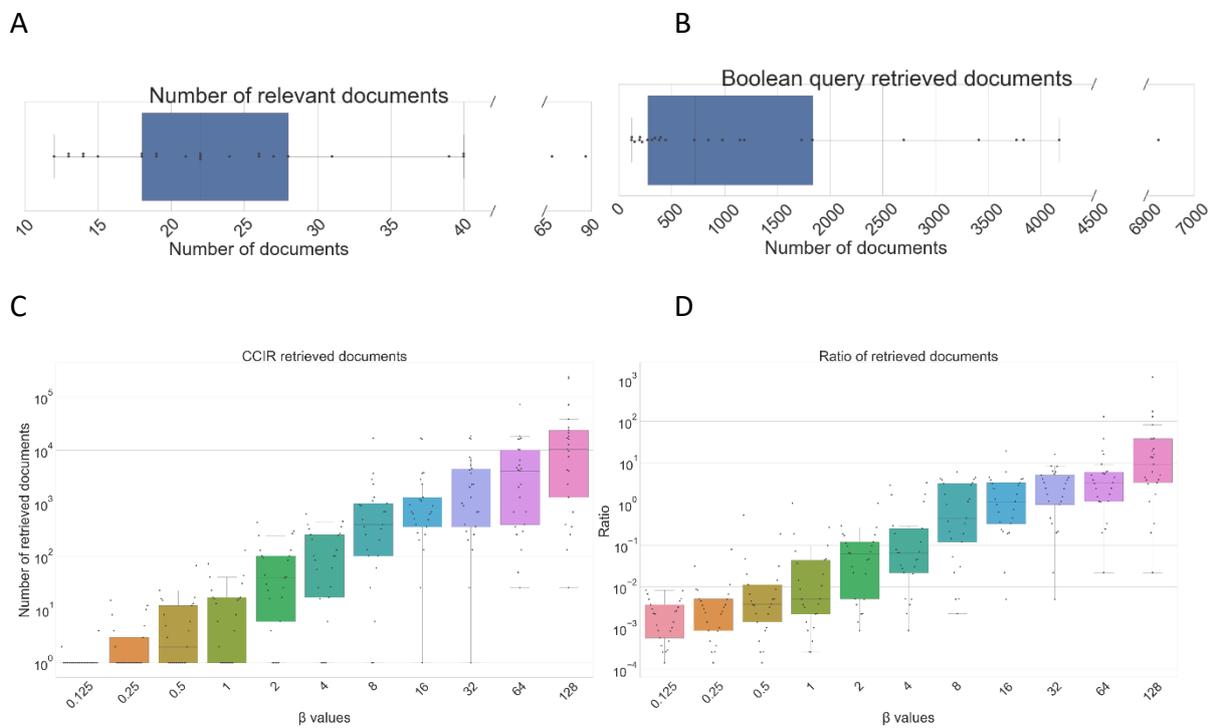

**Figure 4. Documents sets sizes. A**: Relevant documents sets sizes. Each data point is a SR, and the X axis is the size of the relevant documents set. **B**: Boolean query retrieved documents sets sizes. Each data point is a SR, and the X axis is the size of the Boolean query retrieved documents set. **C and D:** Each data point is a SR, the X axis is the β group, and the Y axis is the respective metric of that β group for that SR. **C:** CCIR retrieved documents sets sizes. **D:** Ratio of retrieved documents. Calculated as the CCIR retrieved documents set size divided by the Boolean query retrieved documents set size.

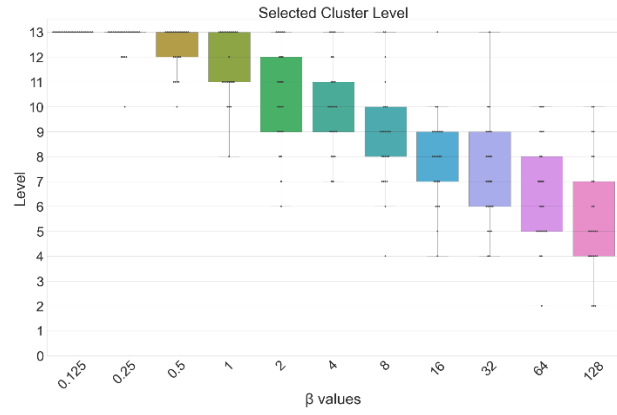

**Figure 5. Tree-level of the retrieved clusters.** Each data point is a SR, the X axis is the β group, and the Y axis is the level of the cluster selected by that greedy algorithm for that SR. Level 0 is the set of all documents in the citation network.

SR59

Optimal cluster                    Parent cluster

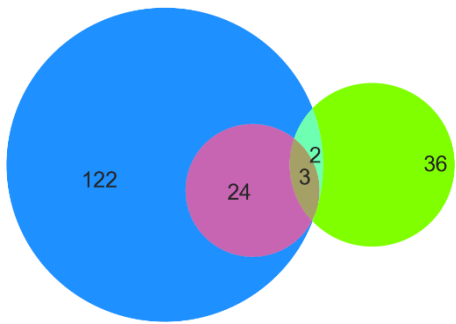 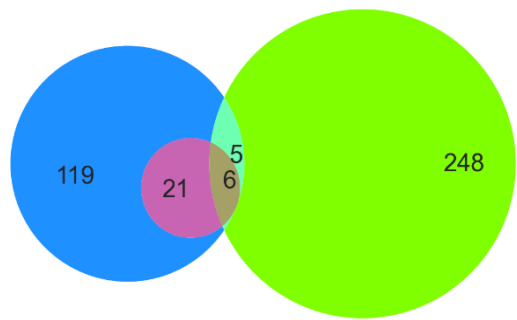

SR47

Optimal cluster                    Parent cluster

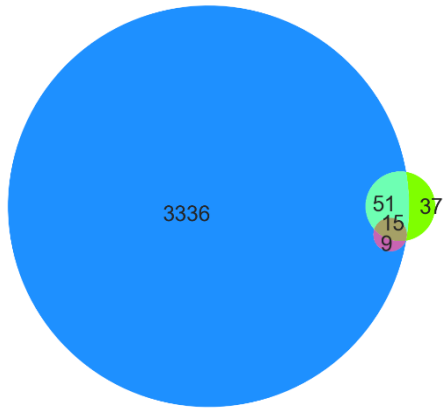 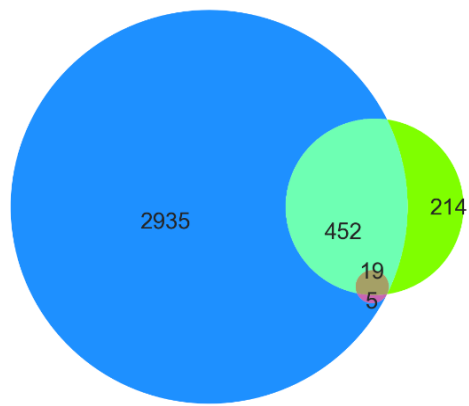

SR80

Optimal cluster                    Parent cluster

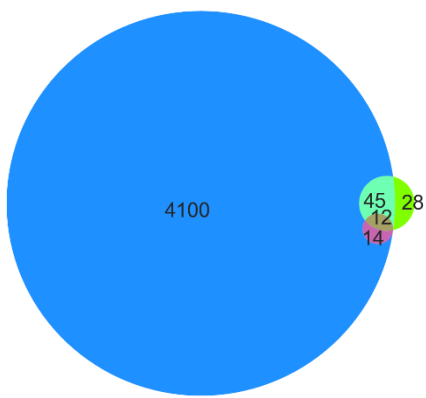 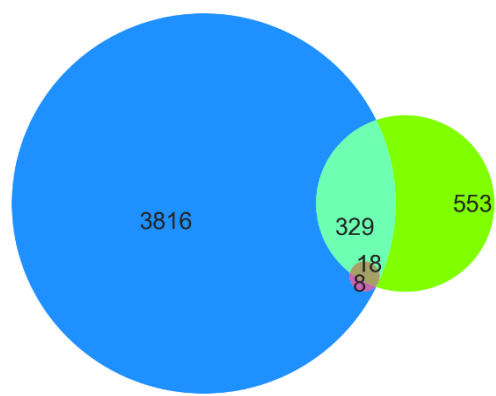

**Figure 6. Venn diagram of the intersections.** Blue: Boolean query retrieved documents set, Green: CCIR retrieved documents set, Red: Relevant documents set.

## Tables

**Table 1. Quantitative data of the SRs in the qualitative analysis.** These are the SRs selected for qualitative analysis (SR59, SR47 and SR80). The F-score values of β = 4 were the ones used to select the SRs. The optimal cluster was selected for its good precision and recall, and the parent clusters because it was the parent cluster of the optimal algorithm (see methods, section 3.5).

| Set of documents | Metric | SR59 | SR47 | SR80 |
|---|---|---|---|---|
| β = 4 | CCIR F-score | 0.15 | 0.52 | 0.41 |
| | Boolean query F-score | 0.79 | 0.11 | 0.1 |
| | F-scores difference | -0.64 | 0.42 | 0.31 |
| Boolean query | Retrieved documents set size | 151 | 3411 | 4171 |
| | Relevant retrieved documents set size | 27 | 24 | 26 |
| | Precision | 0.18 | 0.01 | 0.01 |
| Optimal cluster | CCIR β value | 1 | 2 | 2 |
| | Retrieved documents set size | 41 | 103 | 85 |
| | Relevant retrieved documents set size | 3 | 15 | 12 |
| | Precision | 0.07 | 0.15 | 0.14 |
| | Recall | 0.11 | 0.62 | 0.46 |
| | Intersection set size | 5 | 66 | 57 |
| | Intersection proportion of CCIR | 0.12 | 0.64 | 0.67 |
| Parent cluster | CCIR β value | 4 | 16 | 8 |
| | Retrieved documents set size | 259 | 685 | 900 |
| | Relevant retrieved documents set size | 6 | 19 | 18 |
| | Precision | 0.02 | 0.03 | 0.02 |
| | Recall | 0.22 | 0.79 | 0.69 |
| | Intersection set size | 11 | 471 | 347 |
| | Intersection proportion of CCIR | 0.04 | 0.69 | 0.39 |

**Table 2. Characterization of the SRs.** These are the SRs selected for qualitative analysis (SR59, SR47 and SR80). Goal: The question that the authors of the SR want to answer. Needs: The nature of the documents that the authors need to retrieve to achieve the goal. Boolean query components: The components of which the Boolean query consist. The details on the construction of the Boolean query components are in the supplementary material, Figures S1, S2 and S3.

|  | SR59 | SR47 | SR80 |
|---|---|---|---|
| Title | Retinoic acid post consolidation therapy for high-risk neuroblastoma patients treated with autologous hematopoietic stem cell transplantation | Surgery for the resolution of symptoms in malignant bowel obstruction in advanced gynaecological and gastrointestinal cancer | Rituximab for rheumatoid arthritis (Review) |
| Goal | To determine if **retinoic acid** helps **neuroblastoma** patients recuperate from chemotherapy and bone marrow transplants. | To assess the efficacy of **surgery** for **intestinal obstruction** due to advanced **gynaecological and gastrointestinal cancer**. | To evaluate the benefits and harms of **Rituximab** for the treatment of **Rheumatoid Arthritis**. |
| Needs | **Randomized controlled trials** that evaluate if retinoic acid helps neuroblastoma patients recuperate from bone marrow transplants by **comparing retinoic acid treated patients to untreated patients**. | Documents that mention the **evolution** of patients after **surgeries** to treat **intestinal obstruction** due to advanced **gynaecological and gastrointestinal cancer**. | Studies that compare the outcomes of treatments with **Rituximab** with placebo or other Disease-modifying antirheumatic drugs (**DMARD**). |
| Boolean query components | Retinoic acid AND Neuroblastoma AND Randomized Controlled Trials and Controlled Clinical Trials | Gynecological or gastrointestinal cancer AND Intestinal obstruction AND Surgery | Rheumatoid Arthritis AND Disease-modifying antirheumatic drugs AND Randomized Controlled Trials and Controlled Clinical Trials |

**Table 3. Topic of the sets of documents of the SRs.** These are the SRs selected for qualitative analysis (SR59, SR47 and SR80). We obtained these topics by analyzing the most common noun-phrases in the titles of the retrieved documents. The details on the construction of the topics are in the supplementary material, Tables S3, S4 and S5.

| ID | Set of documents | Topic of the set | Topic of all sets |
|---|---|---|---|
| SR59 | Boolean query | Retinoic Acid for neuroblastoma | Treatments of neuroblastoma |
| | Optimal cluster | Marrow transplant for neuroblastoma. | |
| | Parent cluster | 131I-mibg for neuroblastoma. | |
| SR47 | Boolean query | Disperse topic, most common noun-phrases are too infrequent | Treatment of bowel obstructions in cancer |
| | Optimal cluster | Management of bowel obstructions in cancer, includes non-surgery alternatives | |
| | Parent cluster | More techniques for managing bowel obstructions including emergencies bridge as surgery and self-expandable metal stent. | |
| SR80 | Boolean query | Treat rheumatoid arthritis with several DMARDs | Treatment of rheumatoid arthritis with DMARDs |
| | Optimal cluster | Treat rheumatoid arthritis with few DMARDs | |
| | Parent cluster | Treat rheumatoid arthritis with several DMARDs (including certolizumab pegol) | |

## Supplementary material

### Figures

```
(
        2004/01/01:2014/09/30[Date - Publication]
)
AND
(
        (
                neuroblastoma[Mesh Terms:noexp]
                OR neuroblastomas[Mesh Terms:noexp]
                OR neuroblast*[Mesh Terms:noexp]
                OR ganglioneuroblastoma[Mesh Terms:noexp]
                OR ganglioneuroblastomas[Mesh Terms:noexp]
                OR ganglioneuroblast*[Mesh Terms:noexp]
                OR neuroepithelioma[Mesh Terms:noexp]
                OR neuroepitheliomas[Mesh Terms:noexp]
                OR neuroepitheliom*[Mesh Terms:noexp]
                OR esthesioneuroblastoma[Mesh Terms:noexp]
                OR esthesioneuroblastomas[Mesh Terms:noexp]
                OR esthesioneuroblastom*[Mesh Terms:noexp]
                OR schwannian[Mesh Terms:noexp]
        )
        OR
        (
                neuroblastoma[Text Word]
                OR neuroblastomas[Text Word]
                OR neuroblast*[Text Word]
                OR ganglioneuroblastoma[Text Word]
                OR ganglioneuroblastomas[Text Word]
                OR ganglioneuroblast*[Text Word]
                OR neuroepithelioma[Text Word]
                OR neuroepitheliomas[Text Word]
                OR neuroepitheliom*[Text Word]
                OR esthesioneuroblastoma[Text Word]
                OR esthesioneuroblastomas[Text Word]
                OR esthesioneuroblastom*[Text Word]
                OR schwannian[Text Word]
        )
)
AND
(
        (
                retinoic acid[Mesh Terms:noexp]
                OR retinoic acids[Mesh Terms:noexp]
                OR Retinoid*[Mesh Terms:noexp]
                OR Retinoid[Mesh Terms:noexp]
                OR Retinoids[Mesh Terms:noexp]
                OR tretinoin[Mesh Terms:noexp]
                OR Vitamin A Acid[Mesh Terms:noexp]
                OR Acid, Vitamin A[Mesh Terms:noexp]
                OR trans-Retinoic Acid[Mesh Terms:noexp]
                OR Acid, trans-Retinoic[Mesh Terms:noexp]
                OR trans Retinoic Acid[Mesh Terms:noexp]
                OR all-trans-Retinoic Acid[Mesh Terms:noexp]
                OR Acid, all-trans-Retinoic[Mesh Terms:noexp]
                OR all trans Retinoic Acid[Mesh Terms:noexp]
                OR beta-all-trans-Retinoic Acid[Mesh Terms:noexp]
                OR beta all trans Retinoic Acid[Mesh Terms:noexp]
                OR 3-cis-RA[Mesh Terms:noexp]
                OR 13-cis-retinoic acid[Mesh Terms:noexp]
                OR 4759-48-2[Mesh Terms:noexp]
                OR Retin-A[Mesh Terms:noexp]
                OR Retin A[Mesh Terms:noexp]
                OR Vesanoid[Mesh Terms:noexp]
                OR isotretinoin[Mesh Terms:noexp]
                OR ATRA[Mesh Terms:noexp]
                OR Accutane[Mesh Terms:noexp]
                OR Airol[Mesh Terms:noexp]
                OR Dermairol[Mesh Terms:noexp]
```

```
                )
                OR
                (
                        retinoic acid[Text Word]
                        OR retinoic acids[Text Word]
                        OR Retinoid*[Text Word]
                        OR Retinoid[Text Word]
                        OR Retinoids[Text Word]
                        OR tretinoin[Text Word]
                        OR Vitamin A Acid[Text Word]
                        OR Acid, Vitamin A[Text Word]
                        OR trans-Retinoic Acid[Text Word]
                        OR Acid, trans-Retinoic[Text Word]
                        OR trans Retinoic Acid[Text Word]
                        OR all-trans-Retinoic Acid[Text Word]
                        OR Acid, all-trans-Retinoic[Text Word]
                        OR all trans Retinoic Acid[Text Word]
                        OR beta-all-trans-Retinoic Acid[Text Word]
                        OR beta all trans Retinoic Acid[Text Word]
                        OR 3-cis-RA[Text Word]
                        OR 13-cis-retinoic acid[Text Word]
                        OR 4759-48-2[Text Word]
                        OR Retin-A[Text Word]
                        OR Retin A[Text Word]
                        OR Vesanoid[Text Word]
                        OR isotretinoin[Text Word]
                        OR ATRA[Text Word]
                        OR Accutane[Text Word]
                        OR Airol[Text Word]
                        OR Dermairol[Text Word]
                )
)
AND
(
        (
                randomized controlled trial[Publication Type])
                OR (controlled clinical trial[Publication Type])
                OR (randomized[Title/Abstract])
                OR (placebo[Title/Abstract])
                OR (drug therapy[sh])
                OR (randomly[Title/Abstract])
                OR (trial[Title/Abstract])
                OR (groups[Title/Abstract])
        )
        AND
        (
                humans[Mesh Terms:noexp]
        )
)
```

**Figure S1. SR59 PubMed Boolean query and its components.** The indentations and line breaks represent the dependencies, but they are unnecessary when submitting to PubMed. The colors are the components. Meaning of the colors: Orange: Time Period; Blue: Retinoic acid; Purple: Neuroblastoma; Red: Randomized Controlled Trials and Controlled Clinical Trials. Components ensemble: Time period AND Retinoic acid AND Neuroblastoma AND Randomized Controlled Trials and Controlled Clinical.

```
(
	2005/01/01:2015/06/29[Date - Publication]
)
AND
(
	(
		Gastrointestinal Neoplasms[Mesh Terms]
		OR Genital Neoplasms, Female[Mesh Terms]
		OR Ovarian Neoplasms[Mesh Terms]
		OR
		(
			(
				(
					gynaecological[All Fields]
					OR gynecological[All Fields]
				)
				OR
				(
					gastrointestin*[All Fields]
					OR gastro-intestin*[All Fields]
					OR intestin*[All Fields]
					OR bowel*[All Fields]
					OR colon*[All Fields]
					OR colorectal[All Fields]
					OR rectal*[All Fields]
					OR stomach*[All Fields]
					OR gastric*[All Fields]
				)
				OR
				(
					ovarian[All Fields]
					OR ovary[All Fields]
					OR ovaries[All Fields]
				)
				OR
				(
					endometrial[All Fields]
					OR endometrium[All Fields]
				)
				OR
				(
					uterine[All Fields]
					OR uterus[All Fields]
					OR vaginal[All Fields]
					OR vulvar[All Fields]
					OR vagina*[All Fields]
				)
				OR
				(
					cervix[All Fields]
					OR cervical[All Fields]
				)
			)
			AND
			(
				Neoplasms[Mesh Terms:noexp]
				OR
				(
					neoplasm*[All Fields]
					OR cancer*[All Fields]
					OR carcinoma*[All Fields]
					OR tumor*[All Fields]
					OR tumour*[All Fields]
				)
			)
		)
```

```
                )
        AND
        (
                (
                        Intestinal Obstruction[Mesh Terms]
                        OR
                        (
                                (
                                        bowel*[All Fields]
                                        OR intestin*[All Fields]
                                        OR gastrointestin*[All Fields]
                                        OR gastro-intestin*[All Fields]
                                        OR colon*[All Fields]
                                        OR colorect*[All Fields]
                                        OR retrosigmoid*[All Fields]
                                )
                                AND
                                (
                                        obstruct*[All Fields]
                                        OR blockage[All Fields]
                                )
                        )
                )
                AND
                (
                        Colorectal Surgery[Mesh Terms]
                        OR Surgical Procedures, Elective[Mesh Terms]
                        OR
                        (
                                surgery[All Fields]
                                OR surgical*[All Fields]
                                OR resect*[All Fields]
                        )
                )
        )
)
```

**Figure S2. SR47 PubMed Boolean query and its components.** The indentations and line breaks represent the dependencies, but they are unnecessary when submitting to PubMed. The colors are the components.
Meaning of the colors: Orange: Time Period; Blue: Gynecological or gastrointestinal cancer; Purple: Intestinal obstruction; Red: Surgery. Components ensemble: Time period AND Gynecological or gastrointestinal cancer AND Intestinal obstruction AND Surgery.

```
(
        2004/01/01:2013/12/31[Date - Publication]
)
AND
(
        (
                (
                        arthritis, rheumatoid[Mesh Terms]
                        OR
                        (
                                felty*[Text Word]
                                AND syndrome[Text Word]
                        )
                        OR
                        (
                                caplan*[Text Word]
                                AND syndrome[Text Word]
                        )
                        OR rheumatoid nodule[Text Word]
                        OR
                        (
                                sjogren*[Text Word]
                                AND syndrome[Text Word]
                        )
                        OR
                        (
                                sicca[Text Word]
                                AND syndrome[Text Word]
                        )
                        OR
                        (
                                still*[Text Word]
                                AND disease[Text Word]
                        )
                        OR
                        (
                                bechterew*[Text Word]
                                AND disease[Text Word]
                        )
                        OR
                        (
                                arthritis[Text Word]
                                AND rheumat*[Text Word]
                        )
                )
                AND
                (
                        Antibodies, Monoclonal[Mesh Terms:noexp]
                        OR Immunologic Factors[Mesh Terms:noexp]
                        OR rituximab[Text Word]
                        OR rituxan[Text Word]
                        OR mabthera[Text Word]
                )
        )
        AND
        (
                (
                        clinical trial[Publication Type]
                        OR randomized[Title/Abstract]
                        OR placebo[Title/Abstract]
                        OR drug therapy[MeSH Subheading:noexp]
                        OR clinical trials[Mesh Terms:noexp]
                        OR randomly[Title/Abstract]
                        OR trial[Title]
                        OR groups[Title/Abstract]
                )
                NOT
                (
                        animals[Mesh Terms:noexp]
                        NOT
                        (
```

```
                                    animals[Mesh Terms:noexp]
                                    AND humans[Mesh Terms:noexp]
                                )
                            )
                        )
)
```

**Figure S3. SR80 PubMed Boolean query and its components.** The indentations and line breaks represent the dependencies, but they are unnecessary when submitting to PubMed. The colors are the components. Meaning of the colors: Orange: Time Period; Blue: Rheumatoid Arthritis; Purple: Disease-modifying antirheumatic drugs; Red: Randomized Controlled Trials and Controlled Clinical Trials. Components ensemble: Time Period AND Rheumatoid Arthritis AND Disease-modifying antirheumatic drugs AND Randomized Controlled Trials and Controlled Clinical Trials.

## Tables

**Table S1. Equivalence between OVID symbols and PubMed symbols.** The first column is the OVID symbol, the second column is the PubMed symbol that replaces the OVID symbol, and the third column is the notes about why this PubMed symbol was chosen to replace the OVID symbol.

| OVID | PubMed | Notes |
|---|---|---|
| $ | * | - |
| exp [term]/ | [Mesh Terms] | If a MESH term includes the term "exp", it means that it is an exploding MESH term (https://medlinetranspose.github.io/documentation.html). This means that if a MESH term have the "exp" term, it translates to PubMed as [Mesh Terms], otherwise it translates as [Mesh Terms:noexp]. |
| [term]/ | [Mesh Terms:noexp] | See above. |
| .sh. | [Mesh Terms:noexp] | The library TRANSMUTE translate .sh into [MeSH Subheading], but .sh. means MeSH Subject Headings (https://ospguides.ovid.com/OSPguides/medline.htm). Therefore, I manually replace the [MeSH Subheading] categories translated from .sh categories into [Mesh Terms:noexp] categories.<br>I made sure that the .sh category translates into [Mesh Terms:noexp] and not [Mesh Terms]. I did this by comparing the [Mesh Terms:noexp] and [Mesh Terms] versions of the translations of SR4. The number of retrieved papers with [Mesh Terms:noexp] was closer than [Mesh Terms] to the number of retrieved papers reported by SR4 for the respective lines of the Boolean query. |
| exp *[term]/ | [MeSH Major Topic] | https://libguides.kcl.ac.uk/systematicreview/advanced. The TRANSMUTE library do this translation wrong, so I have to do it manually. |
| *[term]/ | [MeSH Major Topic:noexp] | - |
| .fs. | [MeSH Subheading:noexp] | I believe that the .fs. category means MESH subheading (https://ospguides.ovid.com/OSPguides/medline.htm, https://www.nlm.nih.gov/bsd/disted/meshtutorial/searchingpubmedusingmeshtags/02.html). The TRANSMUTE library translates .fs into [MeSH Subheading:noexp], with is the MESH subheading without explosion. I have found no reference in OVID documentation to whether .fs. explodes or do not, and when I use either [MeSH Subheading:noexp] or [MeSH Subheading] in PUBMED I get the same result (SR6). There is a category in OVID, .xs., that explodes the subheadings, but the TRANSMUTE library translates that category into [All fields], so I guess that the TRANSMUTE library does not recognizes it. Also, the table of MESH subheadings (https://www.ncbi.nlm.nih.gov/books/NBK3827/table/pubmedhelp.T.mesh_subheadings/) doesn't have hierarchies the could be exploded, which suggests that the MESH subheading don't explode . To be consistent with the TRANSMUTE library, I will use the TRANSMUTE translation of .fs. ([MeSH Subheading:noexp]). |
| [term]/[term] | ([Mesh Terms] AND [MeSH Subheading:noexp]) | The term after the slash is a floating subheding (https://libguides.kcl.ac.uk/systematicreview/advanced). |
| .nm. | [All Fields] | .nm. is *name of substance* (https://ospguides.ovid.com/OSPguides/medline.htm), but I could not find an equivalent in PubMed, so i used [All Fields]. I believe that it is okay because names of substances usually don't have homonyms. |
| *# (e. g. *1) | * | PubMed does not recognize the use of numbers for the asterisks. |
| .ti,ab. | [Title/Abstract] | - |
| .ab. | [Title/Abstract] | The category .ab. was translated by the TRANSMUTE library into [All Fields]. According to the documentation of OVID (https://ospguides.ovid.com/OSPguides/medline.htm), .ab means Abstract. PubMed doesn't have a category to search in the abstract, but instead have the category [Title/Abstract] to search in both the title and the abstract (https://www.ncbi.nlm.nih.gov/books/NBK3827/#_pubmedhelp_Search_Field_Descriptions_and_). Therefore, I manually replaced the [All Fields] categories translated from .ab categories into [Title/Abstract] categories. |
| .ti. | [Title] | - |
| .tw. | [Text Word] | - |
| .mp. | [All Fields] | .mp. means multipurpose, which includes many fields (https://ospguides.ovid.com/OSPguides/medline.htm). |
| .af. | [All Fields] | .af. Means all searchable fields (https://ospguides.ovid.com/OSPguides/medline.htm#AF). |
| adj | AND | - |
| adj[#] (e. g. adj[1]) | AND | Some of the queries have proximity operators (e.g. adj3). A webpage (https://utas.libguides.com/SystematicReviews/PhrasesWildcardsProximity) says that PubMed doesn't have proximity operators and another webpage (https://research.library.gsu.edu/c.php?g=115556&p=751933) says that it has. I tried the proximity operator suggested by the latter webpage, and it didn't seem to recognize it as an operator, so I conclude that PubMed doesn't use proximity operators.<br>To solve this issue, instead of proximity operators I will use the operator AND. This change will include all the papers that the proximity operator would had included, but unfortunately it will also include other papers. |
| near | AND | - |
| .pt. | [Publication Type] | - |
| ? | ? | - |
| .ed,dp,yr. | [Date - Publication] | - |
| exp "clinical trial [publication type]"/ | "clinical trial"[publication type] | This line appears in many Boolean queries. It implies that there is a [Mesh Terms] named "clinical trial [publication type]", but this is not recognized by PubMed nor does the terms appears in the MESH terms database.<br>Based on the context of the line and the notes of the authors in some systematic reviews, I believe that authors meant "clinical trial"[publication type], and they copied this line from an standard template. |
| .gc. | NO TRANSLATION | .gc. Grant country (https://ospguides.ovid.com/OSPguides/medline.htm#GC), and PubMed does not have this information available for search (as far as I know). Therefore, I decided to remove the lines with this category.<br>So far, the only SR with .gc. is SR78 (https://www.cochranelibrary.com/cdsr/doi/10.1002/14651858.CD001843.pub5/full). |
| .sh,rn,tw. | [All Fields] | - |
| .sh. | [Mesh Terms:noexp] | Alert: I'm ambiguous about this translation, see below.<br>The library TRANSMUTE translate .sh into [MeSH Subheading], but .sh means MeSH Subject Headings (https://ospguides.ovid.com/OSPguides/medline.htm). Therefore, I manually replace the [MeSH Subheading] categories translated from .sh. categories into [Mesh Terms:noexp] categories.<br>I made sure that the .sh. category translates into [Mesh Terms:noexp] and not [Mesh Terms]. I did this by comparing the [Mesh Terms:noexp] and [Mesh Terms] versions of the translations of SR4. The number of retrieved papers with [Mesh Terms:noexp] was closer than [Mesh Terms] to the number of retrieved papers reported by SR4 for the respective lines of the Boolean query.<br>Correction:<br>I found that SR59 uses the .sh. as a [MeSH Subheading], now I'm not sure if .sh. is [Mesh Terms:noexp] or [MeSH Subheading]. I will avoid SRs that use .sh. in the evaluation so to avoid uncertainty. |

**Table S2. Notes from the translation of the Boolean queries of individual SRs.** The first column is the identity of the SR in the Scells dataset, and the second Colum is the notes from translating the OVID query of that SR into a PubMed query.

| SR | Notes |
|---|---|
| SR3 | There is a part in the Boolean query where 2 clauses are not connected by any operator (emollient$.ti,ab. (skin adj6 (...)).ti,ab.). It seemed to me that (skin adj6 (...)).ti,ab. is another name for creams, so I conclude that the missing operator is OR.<br>I had to remove the asterisk * from oil* and gel* because PubMed needs 4 characters or more for wildcard search. |
| SR4 | There is a term that has a wildcard before the end (8 abnormal menstrua$ bleeding.tw). This can't be translated into PubMed because PubMed doesn't allow asterisks before the end. If I include the asterisk (e.g. abnormal menstrua* bleeding[Text Word]) then PubMed will separate the terms in the asterisk (e.g. abnormal menstrua* AND bleeding[Text Word]). To include this term, I create 2 terms connected by an AND (e.g. abnormal menstrua*[Text Word] AND bleeding[Text Word]). This will retrieve the documents retrieved by the OVID query but unfortunately it will also retrieve another papers.<br>The content of .mp. using [mp=...]. The content of [mp=...] in this Bololean query is similar to the definition of .mp. by OVID (https://ospguides.ovid.com/OSPguides/medline.htm), so I decided to use [mp=...] as a normal .mp. (i.e. translated to [All Fields]).<br>Many times, TRANSMUTE made the following type of mistake: (naproxen.tw. OR NAPROXEN/) is translated to (naproxen.tw. OR NAPROXEN[Mesh Terms:noexp]). This mistake happened when I had OR statements in the same row, even when they were grouped by a parenthesis (e.g. (medical therap$ OR medical treatment$).tw.). I fixed these mistakes manually. |
| SR12 | Because the way the Boolean query is written, it lefts a row with information out (row 25). The query also utilizes the same row two times (row 33).<br>From these observations, I believe there is some mistake in the report of the Boolean query that the authors reported in the paper. For this reason, I will try to reconstitute what I believe was the original query.<br>Row 25 contains info about the chemical component, while row 33 contains info about the trials. Row 37 seems to be an expansion of the concepts in row 33. I see that row 38 joins row 33 and 37 with an OR, which gives weight to my suspicion that row 37 is an expansion of row 33. If this is the case, then there should be a row containing [25 AND 38]. Row 35 is [25 AND 33], so this is the row I am looking for, but it excludes the information in row 37. Therefore, I will include a row [25 AND 38]. The other missing part of the query is any rows that includes (NOT 34), so the final row will be equivalent to [(25 AND 38) NOT 34]. |
| SR169 | Line 5 was (detox* or methadone) in .ti,ab.. I assumed that .ti,ab. are the categories of the line.<br>Line 10 had an incomplete parenthesis, I assumed that the parenthesis encompassed the whole line.<br>Some lines had no categories. I assumed that their category was [All Fields]. |
| SR78 | The Boolean query that was annotated in the Scells dataset is not the same as the Boolean query that was used by the authors of the systematic review (https://www.cochranelibrary.com/cdsr/doi/10.1002/14651858.CD001843.pub5/appendices). I believe that the annotator of the dataset made a mistake. I will be using the Boolean query as reported by the systematic review.<br>The dataset Boolan query says to search until 2006-01-30, but this is wrong. The search in the paper is until 26 February 2014. I will use the latter. |
| SR61 | Line 59 and 109 is used twice, so I duplicated it in the PubMed translation.<br>I removed the date delimitation lines.<br>I gave up on translating this Boolean query. |
| SR59 | The authors use the categories .ms. and .tiab., which are not in the vocabulary of OVID. I believe they mean [MeSH Subheading:noexp] and [Title/Abstract], respectively.<br>The line (randomised controlled trial [pt]) is misspelled. It should read (randomized controlled trial [pt]), otherwise PubMed does not recognize the publication type. I corrected the spelling.<br>Some of the [MeSH Subheading:noexp] terms were not recognized by PubMed. I believe that this is okay because the authors searched for these terms also in the titles and abstract, so they knew that some terms were not valid for the category [MeSH Subheading:noexp]. |
| SR24 | Line 18 had to be fixed, it was missing an OR operator.<br>Line 14 was simplified.<br>There is a [Title/Abstract] that starts with * (*amphetamine[Title/Abstract]) and I don't know that that means. I found no info about it in the documentation (https://ospguides.ovid.com/OSPguides/medline.htm) so I will just remove the *.<br>Lines 28 and 29 make no sense, they don't include the rest of the lines in a logical way. Because of this, I will exclude this SR for now. |
| SR43 | Lines 8 and 9 are excluded from the query. I should analyze the query to understand what was their purpose.<br>Lines 8 and 9 specify that the authors look for clinical trials.<br>Line 7 looks for randomized clinical trials.<br>Line 16 looks for certain type of clinical trials.<br>Line 23 looks for certain type of clinical trials.<br>Line 33 specifies the topic of the clinical trial.<br>The methods of SR43 (https://www.cochranelibrary.com/cdsr/doi/10.1002/14651858.CD005059.pub4/full#CD005059-abs-0003) specify that the SR is looking for clinical trials, therefore, I believe that lines 8 and 9 connect to the query as (AND (8 OR 9)).<br>(operat*[Mesh Terms:noexp]) is recognized by the PubMed website but not by the PubMed API.<br>Some of the rows that use .sh. (comparative study.sh and evaluation studies.sh) are actually publication types. I changed their category to publication types. (evaluation studies[Publication Type]) is not recognized by the PubMed API but it is recognized when used in the PubMed webpage. (evaluation studies[Mesh Terms:noexp]) is not recognized by either. |
| SR47 | Line 15 is not used in the final query. Why?<br>Line 15 is about vaginal cancer.<br>Line 23 is about gastrointestin cirgury.<br>I reviewed the original query of the SR and the Scells dataset is missing a line of the query! Line 24 is present in the original query of the systematic review but not in the dataset, and it contains line 15 and 23.<br>The reason that line 15 was not used in the final query is that the query is incomplete in the dataset. Line 15 is used in the SR. I included the missing line in the PubMed translation. |
| SR51 | Line 29 is used twice, so I duplicated it.<br>Line 36 is used twice, so I duplicated it. |
| SR53 | The query defines .mp.. I believe that the definition the author uses is similar to the general definition of .mp., so I used the standard translation of .mp.. |
| SR115 | I had to discard SR115 because of the size of the translated Boolean query plus the API internet address is too big (11089 characters). Neither the internet browser or the API can manage addresses this long. |
| SR54 | The authors use the term (Succimer/du), but the subheding (du) does not exist. I google Succimier and it is a drug, so the authors must be using du for something related to the drug. It could be Drug Effects (DE) or Drug Therapy (DT). They don't name the word Succimier in their paper, but the paper is about diagnostics, and they include the row Succimer/du in a Boolean parenthesis with (radionuclide image) and (Technetium Tc 99m Dimercaptosuccinic Acid) (a component used in radiology), so I believe that they use the drug for diagnostics and radiology. This doesn't fit the subjedings used for drugs, but it does fit Diagnostic Imaging (DG). Therefore, I will use Diagnostic Imaging for (Succimer/du).<br>The same happens for (Kidney/ri). (ri) is not a subheading, but I believe that the authors refer to kidney radiology. Therefore, I will use Diagnostic Imaging (DG). |
| SR69 | Some of the words in line 3 have no category, so I give them the category that is used by the other words in the row, [Title/Abstract].<br>One row has the concept (work'*[Text Word]), which is interpret as (work*[Text Word]) by PubMed. Therefore, I replaced (work'*[Text Word]) with (work*[Text Word]). Also, the (') can produce issues in the API, so I better do the replacement. |
| SR80 | I had to duplicate line 27 because it is referred to 2 times |
| SR151 | It has many concepts without categories, I will use the category [All Fields] for these concepts.<br>The query uses [mh], which I will translate as [Mesh Terms] (I tested the terms with [mh] as [Mesh Terms] in PubMed and the words are recognized as mesh terms).<br>There is a term (drug therapy[sh]) which I'm sure it means (drug therapy[MeSH Subheading:noexp]) because I have seen it in the other queries. |
| SR130 | The starting date of the query says 2010-12-31 in the test set but it says 2011-01-01 in the paper. I will use 2011-01-01. |
| SR119 | The test set says from 2006-12-31 to 2007-01-20 but the SR says from 2007 to 7 February 2013. I will use the latter. |
| SR62 | Some of the rows that use .sh (comparative study.sh and evaluation studies.sh) are actually publication types. I changed their category to publication types. (evaluation studies[Publication Type]) is not recognized by the PubMed API but it is recognized when used in the PubMed webpage. (evaluation studies[Mesh Terms:noexp]) is not recognized by either. |
| SR105 | (et?omidat*[All Fields]), (r?26?490[All Fields]), and (radenar?on[All Fields]) were not recognized by the by the PubMed API or by the PubMed webpage. I decided to change<br>(et?omidat*[All Fields]) to (etomidate[All Fields]) and (r?26?490[All Fields]) to (r26490[All Fields]) because they refer to etomidate, which i believe is what the author is looking for with this term. I didn't modified (radenar?on[All Fields]).<br>Removed (ill$) (translated into ill*[All Fields]) because PubMed need words to have at least 4 characters to use the asterisk. |

**Table S3. Noun-phrases for SR59.** The noun-phrases were extracted from the titles of the documents.

| Boolean Query 151 titles in total | | Optimal cluster 41 titles in total | | Parent cluster 259 titles in total | |
|---|---|---|---|---|---|
| Noun-phrase | N. of Titles | Noun-phrase | N. of Titles | Noun-phrase | N. of Titles |
| retinoic acid | 25 | risk neuroblastoma | 15 | risk neuroblastoma | 45 |
| neuroblastoma cell | 21 | child | 10 | child | 35 |
| differentiation | 16 | dose chemotherapy | 10 | patient | 34 |
| neuroblastoma | 15 | patient | 5 | neuroblastoma | 32 |
| cis | 11 | stage | 5 | mibg | 24 |
| apoptosis | 11 | tandem | 5 | stage | 19 |
| child | 10 | autologous stem cell rescue | 4 | treatment | 15 |
| risk neuroblastoma | 8 | late effect | 4 | 131i | 15 |
| tran retinoic acid | 7 | autologous stem cell transplantation | 4 | metaiodobenzylguanidine | 15 |
| treatment | 6 | year | 4 | year | 14 |
| retinoid | 5 | treatment | 4 | phase | 13 |
| stage | 5 | age | 3 | dose chemotherapy | 13 |
| neuroblastoma cell line | 5 | trial | 2 | result | 11 |
| human neuroblastoma cell | 4 | autologous stem | 2 | age | 11 |
| cell | 4 | megatherapy | 2 | outcome | 9 |
| combination | 4 | result | 2 | report | 9 |
| trial | 4 | report | 2 | combination | 8 |
| fenretinide | 4 | outcome | 2 | comparison | 8 |
| induction | 4 | feasibility | 2 | risk | 8 |
| cis retinoic acid | 4 | immunotherapy | 2 | irinotecan | 8 |

**Table S4. Noun-phrases for SR47.** The noun-phrases were extracted from the titles of the documents.

| Boolean Query 3411 titles in total | | Optimal cluster 103 titles in total | | Parent cluster 685 titles in total | |
|---|---|---|---|---|---|
| Noun-Phrase | N. of Titles | Noun-Phrase | N. of Titles | Noun-Phrase | N. of Titles |
| patient | 395 | malignant bowel obstruction | 37 | self | 99 |
| case report | 207 | patient | 21 | patient | 91 |
| case | 203 | bowel obstruction | 20 | surgery | 69 |
| self | 121 | octreotide | 12 | management | 54 |
| management | 114 | management | 10 | malignant bowel obstruction | 47 |
| colorectal cancer | 112 | treatment | 9 | bridge | 47 |
| surgery | 110 | palliative surgery | 6 | malignant colorectal obstruction | 45 |
| review | 109 | surgical management | 6 | colorectal cancer | 42 |
| treatment | 101 | outcome | 5 | treatment | 41 |
| report | 100 | advanced ovarian cancer | 5 | malignant colonic obstruction | 41 |
| literature | 88 | predictor | 5 | outcome | 38 |
| intussusception | 77 | peritoneal carcinomatosis | 5 | metallic stent | 35 |
| outcome | 74 | cancer | 5 | comparison | 32 |
| bowel obstruction | 68 | efficacy | 4 | bowel obstruction | 31 |
| year | 63 | stage | 4 | metal stent | 30 |
| analysis | 63 | ovarian cancer | 4 | stent | 28 |
| comparison | 62 | palliative care | 4 | emergency surgery | 26 |
| adult | 59 | intestinal obstruction | 4 | analysis | 25 |
| complication | 58 | palliation | 4 | palliation | 24 |
| chemotherapy | 52 | inoperable bowel obstruction | 4 | expandable metal stent | 24 |

**Table S5. Noun-phrases for SR80.** The noun-phrases were extracted from the titles of the documents.

| Boolean Query 4171 titles in total | | Optimal cluster 85 titles in total | | Parent cluster 900 titles in total | |
|---|---|---|---|---|---|
| Noun-Phrase | N. of Titles | Noun-Phrase | N. of Titles | Noun-Phrase | N. of Titles |
| rheumatoid arthritis | 1362 | rheumatoid arthritis | 46 | rheumatoid arthritis | 501 |
| patient | 1037 | rituximab | 34 | patient | 223 |
| treatment | 518 | patient | 30 | treatment | 158 |
| infliximab | 429 | safety | 16 | safety | 99 |
| rituximab | 255 | efficacy | 10 | tocilizumab | 90 |
| therapy | 198 | treatment | 9 | efficacy | 87 |
| rheumatoid arthritis patient | 180 | result | 8 | methotrexate | 77 |
| adalimumab | 176 | methotrexate | 8 | trial | 55 |
| disease | 170 | use | 6 | rituximab | 54 |
| efficacy | 151 | active rheumatoid arthritis | 5 | active rheumatoid arthritis | 50 |
| safety | 135 | combination | 5 | analysis | 50 |
| effect | 134 | inadequate response | 5 | result | 43 |
| tnf | 127 | therapy | 5 | golimumab | 41 |
| methotrexate | 109 | rheumatoid arthritis patient | 5 | inadequate response | 40 |
| result | 105 | systematic review | 4 | use | 37 |
| case | 101 | recommendation | 4 | year | 35 |
| use | 99 | analysis | 4 | meta | 35 |
| response | 98 | study | 3 | rheumatoid arthritis patient | 33 |
| syndrome | 92 | phase | 3 | placebo | 33 |
| tocilizumab | 85 | clinical outcome | 3 | adalimumab | 32 |